\documentclass[12pt]{article}
\usepackage{setspace}
\usepackage{amsmath,amssymb,mathrsfs}
\usepackage{color}
\usepackage{hyperref}
\usepackage{graphicx,epsfig}
\numberwithin{equation}{section}
\newcommand{\beq}{\begin{equation}}
\newcommand{\eeq}[1]{\label{#1}\end{equation}}
\newcommand{\bea}{\begin{eqnarray}}
\newcommand{\eea}[1]{\label{#1}\end{eqnarray}}

\begin{document}

\begin{titlepage}

\begin{flushright}
\par\end{flushright}
\vskip 0.5cm
\begin{center}
\textbf{\LARGE \bf 
BV Formalism and Partition Functions}\\
\vskip 5mm
\end{center}


\begin{center}

\large 
{\bf 
 \large {\bf P.~A.~Grassi}$^{~a,b,c,}$\footnote{pietro.grassi@uniupo.it} 
 and O. Hulik$^{~d,}$\footnote{ondra.hulik@gmail.com}, 

 \vskip .5cm 
\small

 {$^{(a)}$
 \it DiSIT,} 
 {\it Universit\`a del Piemonte Orientale,} 
 {\it viale T.~Michel, 11, 15121 Alessandria, Italy}
 }
 \small
 
 {$^{(b)}$
 \it Theoretical Physics Department, CERN, 1211 Geneva 23, Switzerland} 
\small
 
 {$^{(c)}$
 \it INFN,} 
 {\it Sez. Torino,} 
 {\it via P. Giuria, 1, 10125, Torino, Italy}
\small

{$^{(d)}$ \it Institute for Mathematics 
 Ruprecht-Karls-Universitat Heidelberg,
 69120 Heidelberg, Germany}

\end{center}
\vspace{1cm}

\vskip  0.2cm
\begin{abstract}

{The BV formalism is a well-established method for analyzing symmetries and quantization of field theories. In this paper we use the BV formalism to derive partition functions of gauge invariant operators up to equations of motions and their redundancies of selected theories. We discuss various interpretations of the results, some dualities and relation to first quantized models.}

\end{abstract}
\vfill{}
\vspace{1.5cm}
\end{titlepage}
\newpage\setcounter{footnote}{0}


\section{Introduction}

Some time ago, interesting analysis have been put forth by A. Connes and M. Dubois-Violette and collaborators in a series of mathematical papers 
\cite{Connes:2002ya,Connes:2003zi,Berger:2002wz}. They provided a "background independent" analysis of interacting gauge theories based on the so-called ``Yang-Mills" algebras taking into account the Euler-Lagrangian equations and their redundancies. In \cite{Connes:2002ya,Connes:2003zi,Berger:2002wz} the complete {\it partition function}
\footnote{In the present work, we use the term partition function to denote, what mathematically more precisely, is known as {\it Hilbert-Poincar\'e series} (see \cite{Berkovits:2005hy}) where different degrees of freedom are counted with a different weights of a fugacity $t$.} for Yang-Mills and self-dual Yang-Mills are obtained by means of the algebraic techniques developed therein. 
Those works have been further analyzed in a series of papers \cite{Movshev:2003ib,Movshev:2004aw,herscovich2010hochschildcyclichomologyyangmills,
Herscovich_2012,Herscovich_2015} \footnote{For a nice review of index techniques see \cite{Gadde:2020yah}.}and extended to the supersymmetric cases. Moreover it is also point out how the homogeneous algebras emerge as limits of algebras appearing in the gauge theory of D-branes and open string theory as shown by N. Nekrasov. 

Given such the partition function for YM and self-dual YM, one can easily construct all possible gauge invariant operators using the standard techniques as illustrated by A. Hanany {\it et al.} 
\cite{Feng:2007ur,Benvenuti:2006qr,Hanany:2008kn,Gray:2008yu, Hanany:2010zz,Cremonesi:2014kwa}. There a long list of applications and computations of partition functions has been derived and discussed. Recently, 
a surge of interest emerged in computation of the space of gauge invariant operators in Effective Field theories 
\cite{Lehman:2015via,Henning:2015daa,Lehman:2015coa,Henning:2017fpj, Graf:2020yxt,Delgado:2022bho} all of which are based on the Hilbert series techniques listing all operators with multiple fields and their derivatives. 

In the present work, we start from a different perspective. We notice that the partition function described in the quoted papers can be understood from a different point of view, namely from BV formalism for gauge theories \cite{HT-book, Barnich:2000zw}.  
The BV formalism is a very useful tool to study the complete structure based on symmetries and equation of motions of a given quantum field theory model. It provides a compact, efficient and elegant way to deal with gauge symmetries, rigid symmetries, open algebras, field-dependent symmetries together with the equations of motion. To each physical field of theory and for each ghost or ghost-for-ghost there is a corresponding antifield (in the forthcoming sections, we will show that it is not necessarily true, see also \cite{Grassi:1999nb,Barnich:2018nqa}), then the gauge symmetries and the equations of motion, together with their redundancies, are translated into the BV-BRST symmetry for the fields and antifields and everything is nicely encoded into an action (BV action), a symplectic form or, and into an anti-bracket.

The quantum numbers of the antifields are fixed in terms of the corresponding fields and the BV-BRST variation of the antifield implements at the cohomological level the equations of motion and their 
redundancies (see \cite{Barnich:2000zw}). Moreover, we assign  suitable charges (scale) to each field and to each ghost field, we fix the charges of the antifields by means of the BV-BRST symmetry and we derive the partition function by counting the independent degrees of freedom. Performing such a counting for Yang-Mills (or the Maxwell for simplicity), one immediately sees the correspondence with the partition function computed by the homogeneous algebra in \cite{Connes:2002ya}. 

Note that the antifields are not quantum propagating fields and therefore the computation of the partition function listed in the forthcoming section cannot be obtained by quantum field theory path integral, but rather by particle/superparticle models or even superstrings. In that respect, we quote the recent work \cite{Boffo:2024lwd} where N=2 superparticle in D=4, where 
the free model allows us to construct explicitly the partition function revealing indeed that it was describing, in a fixed sector of $R$-symmetry, Yang-Mills theory with the complete BV structure. Older works on the subject are based on pure spinors \cite{Berkovits:2005hy,Grassi:2005jz} where the partition function is discussed and constructed from the algebraic analysis of the pure spinor spaces. In those cases when a worldline/worldsheet computation is possible, we find agreement with the computation in \cite{Connes:2002ya}. 

Once we established such a procedure to determine the partition function, we have a powerful instrument to compute it for several models. In addition, in those cases where self-dual $p$-forms are present -- for which no BV-BRST formalism is known -- we can be reversed engineered from the partition function discovering the BV spectrum of fields and antifields and their BV-BRST transformation rules. A very interesting consequence in those cases is that the BV formalism does not seem to have the canonical structure: the pairing between fields and antifield is missing, the symplectic structure is missing and, finally, the BV bracket is also absent. Nonetheless, this is perfectly consistent since the antifield of the gauge field, in the case of Maxwell/Yang-Mills is needed to implement the corresponding equations of motion, but those are already imposed by the self-duality condition. Therefore, a part of the antifield spectrum is unnecessary. This original observation has guided us to construct several examples of partition function with self-dual $p$-forms, to check their on-shell degrees of freedom and to obtain the set of gauge invariant operators. On the other hand, we discover a new antifield implementing the self-duality at the level of BV-BRST cohomology. At the moment we have not explored to consequence of this new BV spectrum, but this is certainly consistent with the common lore on the absence of a covariant, polynomial, finite-number of fields action for self-dual forms (see \cite{Siegel:1983es,Henneaux:1988gg}). 

In order to check that the proposed partition function are correctly described the proposed models, we provide a second interpretation leading to the 
identification of the physical degrees of freedom. This is verified for each model described in the paper. Finally, 
where this construction is valid, under which assumptions, are working in an asymptotic free theory, we are neglecting the quantum corrections and even the interactions \cite{Budzik:2023xbr}.

In sec. 2, we review the main ingredients of the partition function construction by means of BV correspondence. In sec. 3, we study 
several examples of bosonic models with/without self-dual $p$-forms. It is shown how to identify the physical degrees of freedom and how to compute the gauge invariant operators by means of the plethystic polynomials. In addition, it is shown that in presence of non-minimal sector of the BV formalism, the partition function does not change.  In subsection. 3.2 the self-dual Maxwell/YM is studied and the reverse engineering to the BV formalism is obtained. In sec. 3.4 also Einstein gravity is taken into account. In sec. 4, supersymmetric models are studied along the same lines. Also the supersymmetric self-dual cases are considered providing both the partition function and the BV spectrum. In the case of supersymmetric models, the auxiliary fields are taken into account and it is shown how to they emerge in our framework. In sec. 5 the duality (BV duality) is shown and finally in sec.6 a CFT perspective is provided. 
Some appendices conclude the paper. 

\section{From BV-BRST Formalism to Partition Functions}

In this section we outline the main technique of this paper. We recall some basics facts about the BV formalism.  We denote by $\phi_A$ a set of fields and the multi-index $A$ stands for any quantum number needed to identify the field spanning a linear representation of the symmetry group. In addition, the 
BV-BRST symmetry for $\phi_A$ which is parametrized by ghost $C_{A'}$ (again we denote by $A'$ the set of quantum numbers for the ghot fields). The BV-BRST transformation of the ghosts are parametrized by ghost-for-ghost  $C_{A''}$ (again we denote by $A''$ the set of quantum numbers for the ghost-for-ghost fields). Here we provide only a very schematically description, for further details, we refer to the literature \cite{HT-book, Barnich:2000zw}
\begin{eqnarray}
\label{BVA}
s\, \phi_A = \mathcal{F}_{A}(C_B, \phi_A) \,, ~~~
s\, C_{A'} = \mathcal{G}_{A'}(C_{B''}, C_{A'} , \phi_A) \,, ~~~
s\, C_{A''} = \mathcal{H}_{A''}(C_{B'''}, C_{B''}, C_{A'} , \phi_A) \,, ~~\dots
\end{eqnarray} 
where $\mathcal{F}_{A}, \mathcal{G}_{A'}\, \dots$ are functional-differential operators describing rigid or gauge symmetries. 
The closure of the algebra for the algebra is translated into the nilpontency of the BV-BRST differential $s$ and follows from the 
Jabobi identities, equations of motion and their relationships. Contextually to each field $\phi_A, C_{A'}, C_{A''}, \dots$ one introduces a corresponding antifield 
$\phi^{\star A}, C^{\star A'}, C^{\star A''}, \dots $ and they are paired by the symplectic form 
\begin{eqnarray}
    \label{BVB}
    \Omega = \int d^dx \left( \delta \phi_A \delta \phi^{\star A}+ \delta C_{A'} \delta C^{\star A'} +
    \delta C_{A''} \delta C^{\star A''}+ \dots \right) 
\end{eqnarray}
Notice that the antifields carry both a ghost number such that the symplectic form has a global negative ghost number $-1$ 
and the antifield number $+1$. The integration in \eqref{BVB} is on the worldvolume of the theory. In terms of the symplectic form 
one can establish the antibracket  defined as follows 
\begin{eqnarray}
    \label{BVC}
    \left(F, G \right) = \int d^dx \left( \frac{\partial F}{\partial \phi_A}  \frac{\partial G}{\partial \phi^{\star A}} + 
    \frac{\partial F}{\partial C_{A'}}  \frac{\partial G}{\partial C^{\star A'}} + \dots -   F \rightleftharpoons G \right)  
\end{eqnarray}
and there exists a master action $S$, satisfying the classical BV master equations $(S,S) = 0$.\footnote{There is a quantum version of it, but we are not considering this here.} The variation of the fields and of the antifield are obtained by the linearized version of the 
antibracket as follows 
\begin{eqnarray}
    \label{BVD}
    s \phi_A &=& (S, \phi_A)\,, ~~~~~~ s C_{A'} = (S, C_{A'}) \,, ~~~~s C_{A''} = (S, C_{A''}) \,, \dots \nonumber \\
    s \phi^{\star A} &=& (S, \phi^{\star A}) = \frac{\delta S}{\delta \phi_A}\,, ~~~~~~ s C^{\star A'} = (S, C^{\star A'}) \,, ~~~~s C^{\star A''} = (S, C^{\star A'}) \,, \dots    
\end{eqnarray}
where it is easy to observe that the variation of an antifield contains the equations of motion of the corresponding field and 
some covariantization terms. Therefore, as discussed in textbooks \cite{HT-book,Barnich:2000zw}, the equations of motion are implemented at the level of the BV-BRST cohomology under the name of Koszul-Tate differential. The nilpotency of the BV-BRST symmetry extended to the antifield implements the field equations and the relations among them. We have to mention that the case of open algebra implies quadratic or higher powers of antifields $\phi^*$ in the action. The symplectic structure has a twofold effect: firstly, if we assign a given charge to a fields, the antifield necessarily carries a compatible charges, the second effect is the field-antifield duality which is manifest in the partition function (see also \cite{Aisaka:2008vw}). 

The easiest way to assign a charge (which seems to work for several models) is by choosing the 
engineering dimension of the fields $d_\phi$. The corresponding antifield will have 
the complementary dimension $d_{\phi^*} = d - d_\phi$, where $d$ is the dimension of the spacetime. 
The same works for ghost and ghost-for-ghost. Note that the symplectic form respects such assignment because 
the variation of the fields and antifields are integrated over the spacetime. In addition, with this simple assignment  the spacetime derivative $\partial_\mu$ carries charge $+1$. Therefore, we have a simple way to build the partition 
function as follows 
\begin{eqnarray}
    \label{BVE}
    \mathbb{P}(t) = \frac{ \sum_{X}  (-1)^{\epsilon_X} {\rm dim}[X]  t^{d_{X}}
    }{(1-t)^d}
\end{eqnarray}
where the sum is extended to $X = \phi, C, C', \dots, \phi^*, C^*, C^{*'} \dots$ to all fields and antifields of the theory. ${\rm dim}[X]$ is the dimension of the linear space spanned by the field $X$ (for example if $X = \phi_A$, 
${\rm dim}[\phi_A]$ counts the dimension of the vector space spanned by the quantum number $A$ of the fields $\phi_A$. 
The powers $t^{d_{X}}$ take care of the scale of each individual field and $ (-1)^{\epsilon_X}$ takes care of the parity of the fields. We recall that the BV-BRST differential is an antiderivation and thereore at each generation, fields $\phi_A$, ghost $C_{A}$, ghost-for-ghost $C_{A'}$ the parity flips. This is important since the partition function 
takes into account the parity of the set of fields (like in the Euler character in differential geometry). The denominator $(1-t)^d$ takes into account the derivatives of the gauge invariant operators.

The BV formalism is homological in its nature and as such highly non unique. This means that two off-shell inequivalent formulations can have the same cohomology, and therefore the same partition function. Even though in practise it might be hard to identify if two complexes are the same (without calculating the cohomology) upon field redefinition they can only differ by addition of Koszul pairs\footnote{Since we are working with BV formalism and each field has their antifield these are actually quadruplets of the fields.} $\Phi \,, \eta \,, \eta^{\ast} \,, \Phi^{\ast}$   
with the BV-BRST trasformations 
\begin{gather} \label{KT}
        s \Phi = \eta \, \qquad s \eta = 0\,, ~~~~~
        s \Phi^{\ast} = 0 \, \qquad s\,  \eta^{\ast} = \phi^{\ast} 
\end{gather}
from where one can see that the scale of $\Phi,\eta$ must be equal but the cohomological degree opposite. So the contributions in the partition function have to cancel out. This will be shown explicitly in \eqref{SDBAD}.

\section{Non-Supersymmetric Models}

\subsection{1- form Gauge Fields}
We first recall the partition function for $d=4$ Maxwell, we denote by $t$ the scaling power associated to fields 
\begin{eqnarray}
\label{SDA}
\mathbb{P}_{MX}(t) = \frac{1 - 4 t + 4 t^3 - t^4 }{(1-t)^4} = \frac{(1- t^2)}{(1-t)^4} ( 1 + t^2 - 4t)   
\end{eqnarray}
There are several ways to compute this expression, but we refer to 
\cite{Connes:2002ya,Connes:2004xy} where the computation is based on the cubic algebra for YM.

The factor $(1-t^2)$ denotes the on-shell condition, setting $t\rightarrow 1$ put the 
states on-shell. The factor $(1-t)^4$ denotes the derivative on the fields. 
Expanding to the first two orders that expression we have 
\begin{eqnarray}
\label{SDB}
\mathbb{P}_{MX}(t) = 1 - 6 t^2 - 16 t^3 + \mathcal{O}(t^3)
\end{eqnarray}
where the first terms correspond to the constants (present in the cohomology), the term $- 6 t^2$ denote the gauge invariant operators $F_{\mu\nu} = \partial_{[\mu} A_{\nu]}$, and the term $- 16 t^3$, correspond to the gauge invariant operators 
$\partial_\rho F_{\mu\nu}$, which are 24 independent gauge invariant fields. We have however to remove 4 because of the Bianchi's identities and 4 because of the equations of motion, and we are left with 16 DOFs. 

Factoring out the $(1-t^2)/(1-t)^4$, we are left with the polynomial $1 + t^2 - 4t$ 
and in the limit $t\rightarrow 1$, gives $-2$ which are the two degrees of freedom of the photon. Notice that $1+ t^2$ have a ghost/anti-ghost interpretation, while $-4 t$ are the four gauge boson DOFs. 

Note that there is a dual interpretation of the first expression \eqref{SDA}, the polynomial $1 - 4 t + 4 t^3 - t^4$, can be read as counting the DOFs of the 
ghost field $C$, of the gauge field $A_\mu$, of the antifield of the gauge field $A^*_
{\mu\nu\rho}$ (which is a $3$-form) and finally of the antifield of the ghost field $C^*$ (which is a $4$-form). In the present case, the scale $t$ corresponds to the form degree. The BRST transformations relating the various fields are 
\begin{eqnarray}
\label{SDBA}
s C = 0\,,  ~~~~~
s A_\mu = \partial_\mu C\,, ~~~~~~
s A^*_{\mu\nu\rho} = \epsilon_{\mu\nu\rho\sigma } \partial_\tau F^{\sigma\tau} \,, ~~~~~
s C^*_{\mu\nu\rho\sigma} = \partial_{[\mu} A^*_{\nu\rho\sigma]}\,. 
\end{eqnarray}
It is convenient to define $A^{*, \mu} = \epsilon_{\mu\nu\rho\sigma } A^*_{\nu\rho\sigma}$ and $C^{*} = \epsilon_{\mu\nu\rho\sigma } C^*_{\mu\nu\rho\sigma}$ to write the BRST symmetry as
\begin{eqnarray}
\label{SDB-AB}
s C = 0\,,  ~~~~~
s A_\mu = \partial_\mu C\,, ~~~~~~
s A^{*,\mu} = \partial_\tau F^{\mu\tau} \,, ~~~~~
s C^* = \partial_{\mu} A^{*,\mu}\,. 
\end{eqnarray}
The cohomology of this differential operator is given precisely by the partition function \eqref{SDA}. 
Notice that in the cohomology the presence of the antifield $A^*$ is needed to impose the equations of motion for the gauge field (the dynamic is contained in that variation). The antifield $C^*$ is needed in order to taken into account the relations among the Maxwell equations. Finally, notice that 
using the potential $A$ for expressing the field strength $F$ in terms of it, implies that the Bianchi's identities are already solved.

 In paper \cite{Boffo:2024lwd}, a different BV-BRST formalism has been used. 
 Namely, a non-minimal field $\phi$ (it is sometimes identified with the Nakanishi-Lautrup field) and its antifield $\phi^*$ . In particular, the BRST transformations are given by 
\begin{eqnarray}
    \label{SDB-AA}
    &&s C =0\,, ~~~~~s A_\mu = \partial C\,, ~~~~ s \phi = \partial^2 C\,, ~~~~~~
    \nonumber \\
    &&s C^* = \partial^\mu A_\mu^* \,, ~~~~~~
    s A^*_\mu = \partial^\nu F_{\mu\nu} \,, ~~~~
    s \phi^* = -\partial^\mu A_\mu + \phi
\end{eqnarray}
Adopting the charges as above, we see that the scale of $\phi$ and of $\phi^*$ is the same and fixed by the BRST symmetry and then we have 
\begin{equation}
    \label{SDBAD}
    \mathbb{P}_{MX}(t) = \frac{1 - 4 t + t^2 - t^2 + 4 t^3 - t^4}{(1-t)^4} = 
    \frac{(1 - t^2)}{(1- t)^4}{(1 - 4 t + t^2)}
\end{equation}
where the two central terms $t^2 - t^2$, associated to $\phi$ and $\phi^*$, respectively cancel and this gives the same expression as above. 
There is a small detail which is relevant: using the NL field $\phi$,  factor $(1- 4 t + t^2)$ emerges automatically and this is due to the fact that the addenda $1 + t^2$ represent the ghost and the antighost field. The latter becomes dynamical when the gauge fixing is added to the theory. 

In the BV formalism, the addition of a non-minimal sector to the theory in order to implement the gauge fixing does not change the number of gauge invariant operators and the number of physical degrees of freedom. Indeed, the addition of the fields $\phi, \phi^*$ do not alter the partition function, even though it is essential to perform Feynamn diagram computations.


Another way to arrive to the \eqref{SDA} is from the world line model perspective discussed in \cite{Boffo:2024lwd} to which we refer for further details on the model. In that paper the partition function has been constructed for the set of the fields  $\{ X, P, \psi , \gamma, \beta , c \}$ and their conjugates. The BRST symmetry of the model, implemented at the worldline level with the BRST charge $Q$ is 
translated at the targetspace theory by the equation 
\begin{eqnarray}
    \label{SDBAC}
    Q |\,\omega\rangle =  s\,  |\,\omega\rangle 
\end{eqnarray}
where $|\omega\rangle$ are states in the Hilbert space whose coefficients are in the targetspace. \eqref{SDBAC} correctly encodes also the BV fields and the BV-BRST symmetry discussed here.

The charges of the worldline fields are
\begin{center}
\begin{tabular}{|c|c|c|c|c|c|c|}
\hline
Field & $X^m$ & $\psi^m$ & $c$ & $\gamma$ & $\beta$ & 
$P_m$ 
\nonumber \\
\hline
$r$
& 0 & 1 & 0 & 1 & 1  
& 0  
\\
\hline
$s$ 
& 0 & 0 & +1 & +1 & --1  
& 0    \\
\hline
$t$ 
& 0 & -1 & 2 & 0 & -2  
& -1    \\
\hline
 & 1 & $rt^{-1}$ & $st^{2}$ & $sr$ & $\frac{r}{st^2}$ & $t$ \\
 \hline
\end{tabular}

\end{center}
the partition function then reads
\begin{equation}
\mathbb{P}_{N=2}(t,s) = \frac{1}{(1-t)^4} \frac{(1 +  \tfrac{r}{t})^4 (1 +st^2)}{(1 - sr  )(1 - \tfrac{ r  }{st^2})}
\end{equation}
Now, upon expanding this formula around $r=0$ to the second order and specializing $s = -1$ (namely by setting cohomological degree to its conventional value) we arrive to the same function as \eqref{SDA} up to prefactors $(r/t^2)^\alpha$
\begin{eqnarray}
    \label{MAH}
 \mathbb{P}_{N=2}(t, s=-1) &=&     
\frac{(1-t^2)}{(1 - t^2)^4}  
- \left(\frac{r}{t^2}\right) \frac{(1 -t^2) (t^2 - 4 t + 1)}{(1- t)^4} \nonumber \\
&+& \left(\frac{r}{t^2}\right)^2 
\frac{(1-t^2) (t^4 - 4 t^3 + 7 t^2 - 4 t + 1)}{(1-t)^4} + O\Big(\left(\frac{r}{t^2}\right)^3\Big)
\end{eqnarray}
    
Notice that the first 
term in \eqref{MAH} represents a scalar field $\varphi$ and it antifields $\varphi^*$, the second term of the expansion to the Maxwell/YM sector and finally the third term corresponds to 2-form in $D=4$. It is convenient to illustrate also this case before discussing the self-dual cases. 

\subsubsection{2-form in $D=4$}

Since the partition function computed in $N=2$ worldline \eqref{MAH} describes also a $2$-form $B$, we 
illustrates this sector. As is well-known, a $2$-form in $D=4$ has only one DOF. This can be easily checked by observing that the field strength $H = d B$ can be dualized $ \star H = d \varphi$. 
This is a consequence of the Poincar\'e lemma for flat $\mathbb{R}^{1,3}$, of the free equations of motion and Bianchi identities 
\begin{eqnarray}
    \label{MAHA}
    d \star H =0 \Longrightarrow \star H = d\varphi\,, ~~~~~~
    d H =0 \Longrightarrow d\star d \varphi =0\,. 
\end{eqnarray}
Note that solving the equations of motion by dualizing the field, implies that the antifield $B^*$ for the $2$-forms is no longer needed, and a consequence also the ghost and the ghost-for-ghost, but there is small remainder. Indeed we can write the partition function in the following ways 
\begin{eqnarray}
    (1-t^2) \frac{1 - 4t + 7 t^2 - 4 t^3 + t^4}{(1-t)^4} &=&  \frac{(1-t^2) }{(1-t)^4}
   \left( (1 + t^2 + t^4) - 4t(1 + t^2)  + 6 t^2 \right) \nonumber \\
    &=& 
     (1-t^2)   + \frac{(1-t^2) t^2}{(1-t)^4}
\end{eqnarray}
The second expression shows that DOF $(1+ t^2 + t^4)$ are the ghost-for-ghost, $-4 t(1+t^2)$ are the ghost and the antighost and finally $6 t^2$ are the components of the $2$-form in $D=4$. In the limit $t \rightarrow 1$ it gives one, which is the DOF of a scalar. The third expression 
indicates that the first term is the residual zero modes (no derivatives are acting on it since the usual denominator $(1-t^4)$ is missing) from the BV spectrum 
(the ghost-for-ghost zero mode and it antifield), while the second term 
stands for a scalar field (with an additional factor $t^2$, which is remnant of the partition function \eqref{MAH})  which is the dual field for the $2$-form.

\subsection{Self-dual Maxwell}

In \cite{Connes:2004xy}, it is discussed also the partition function for 
self-dual case, and it reads 
\begin{eqnarray}
    \label{SDC}
    \mathbb{P}_{MX_+}(t) = \frac{1 - 4 t + 3 t^2}{(1-t)^4} = \frac{(1- t^2)}{(1-t)^4} \left( 1 - \frac{4t}{1+t}\right)   
\end{eqnarray}
where $MX_+$ stands for self-dual Maxwell. This partition function is computed in 
\cite{Connes:2002ya} based on a 2-homogeneous algebra directly related to the self-duality condition. In that paper, it is shown that due to the properties of the algebra the coefficients in the polynomial in the numerator are fixed, while the numerator stands for the derivatives with gauge invariant operators. 

The interpretation of the partition function is similar to Yang-Mills case, factoring  $(1-t^2)/(1-t)^4$, we are left with $(1 - 4t/(1+t))$ which in the limit gives $-1$ which is the single DOF of self-dual Maxwell (we are assuming Euclidean or $(2,2)$ signature). In the present case, due to the self-duality condition one ghost is removed and the gauge DOFs are halved (the factor $1/(1+t)$). 

How can one reconcile \eqref{SDA}  with \eqref{SDC}? In order to provide a bridge between the two expressions, we notice that there is no power $t^2$ in \eqref{SDA}, but 
we can add and subtract it as follows 
\begin{eqnarray}
\label{SDD}
\mathbb{P}_{MX}(t) = \frac{1 - 4 t + 3 t^2 - 3 t^2 + 4 t^3 - t^4 }{(1-t)^4} = 
\frac{1 - 4 t + 3 t^2}{(1-t)^4} -  t^4 \frac{1 - 4 t^{-1} + 3 t^{-2}}{(1-t)^4} 
\end{eqnarray}
where the inserted and removed $3 t^2$ and we separate $\mathbb{P}_{MX_+}(t)$ from 
the rest which has a "dual" form. By selecting only the self-dual part we are left 
with the first piece of the partition function and we forget about the second term. 
Following the interpretation above, we have the self-dual DOF. However, we can 
also try to provide a BV interpretation of the partition function, by observing that 
term $3 t^2$ might correspond to an antifield (which have negative ghost number), $2$-form (which justify the scale $t^2$) and self-dual (to have $3$ DOFs as listed in 
the partition function). We denote is as $\Sigma^*_+$. 

As above, we can describe the BRST symmetry in the present case. 
We have 
\begin{eqnarray}
    \label{SDE}
    s C =0 \,, ~~~~~~
    s A_\mu = \partial_\mu C\,, ~~~~~
    s \Sigma^*_{\mu\nu, +} = F_{\mu\nu, +}
\end{eqnarray}
which is nilpotent since $F_{\mu\nu, +}$ is BRST invariant. The cohomology of this BRST differential is described by the partition function  $\mathbb{P}_{MX_+}(t)$. Note the fact that 
the self-dual part of the field strength is BRST exact, implies also the equations of motion by means of the Bianchi identities. The role of $A^*$ in the non self-dual case, is superseded by 
$\Sigma^*$. To show this we observe the BRST variation of $A^*$ can be rewritten as follows 
\begin{eqnarray}
    \label{SDEA}
    s A^\star_\mu = \partial^\tau (F_{\tau \mu} + \epsilon_{\mu\tau \rho\sigma} F^{\rho\sigma}) 
\end{eqnarray}
and the second piece is zero because of the Bianchi identities. Using 
\eqref{SDE}, we can replace the r.h.s. of \eqref{SDEA} with 
\begin{eqnarray}
    \label{SDEB}
    s A^\star_\mu = \partial^\tau s \Sigma^*_{\mu\tau, +}
\end{eqnarray}
so we can observe that, up some exact term, we can write
\begin{eqnarray}
    \label{SDEC}
    A^*_\mu = \partial^\tau \Sigma^*_{\mu\tau}
\end{eqnarray}
and therefore the field $A^*_\mu$ becomes superfluous. In the same way, inserting \eqref{SDEC} into the BRST variation of $C^*$, we see that it it vanishes and therefore, $C^*$ is no longer needed and should be removed. This the reason that the fields $A^*_\mu$ is superseded by $\Sigma^*_{\mu\nu}$. 

Essentially, this is also the content of the discussion in \cite{Connes:2004xy} where they show that the self-dual YM algebra is 2-homogeneous and this implies the equations of 
motion. 

\subsection{$2$-form in $D=6$ and its Self-dual Counterpart}

We consider the case of $2$-form in $D=6$. The partition function is 
\begin{eqnarray}
    \label{2FA}
    \mathbb{P}_{2F}(t) &=& \frac{1 - 6 t + 15 t^2 - 15 t^4 + 6 t^5 - t^6}{(1-t)^6} \nonumber \\ 
    &=& \frac{(1- t^2)}{(1-t)^6} \Big( 
    (1+t^2 + t^4) - 6 t(1+ t^2) + 15 t^2 \Big)   
\end{eqnarray}
  which describes a $2$-form. The interpretation of the last factor is always as follows $- 6 t (1+ t^2)$ are the usual ghost for the primary gauge symmetry, $(1+ t^2 + t^4)$ are the ghost-for-ghosts and finally $15 t^2$ are the $B$-field DOF's. Taking the limit $t\rightarrow 1$, it gives the on-shell DOF's which are six. On the other side, one can read from the first expression of \eqref{2FA} the BV content of the model. 

Using the same trick as above, we add and subtract $- 10 t^3 + 10 t^3$ to get the self-dual partition function
\begin{eqnarray}
    \label{2FB}
    \mathbb{P}_{2F}(t) &=& \frac{1 - 6 t + 15 t^2 - 10 t^3 + 10 t^3 - 15 t^4 + 6 t^5 - t^6}{(1-t)^6} \nonumber \\ 
    &=&    
    \frac{1 - 6 t + 15 t^2 - 10 t^3}{(1-t)^6} + 
    t^3 \frac{10 - 15 t + 6 t^2 - t^3}{(1-t)^6}
\end{eqnarray}
where again the first term in the last expression gives the 
partition function of the self-dual $2$-form
\begin{eqnarray}
    \label{2FC}
    \mathbb{P}_{SD-2F}(t) = \frac{1 - 6 t + 15 t^2 - 10 t^3}{(1-t)^6} = \frac{(1-t^2)}{(1-t)^6} \frac{(1 - 5 t + 10 t^2)}{1+t}  
\end{eqnarray}
The last factor in the limit $t\rightarrow 1$, gives the correct DOF's of the self-dual form $B$, with $H = \star H$. Again, notice that the BV antified for $B_{\mu\nu}, C_\mu,C$ are not needed since the self-duality condition put the theory on-shell. 
The BRST symmetry in the present case becomes 
\begin{eqnarray}
    \label{2FD}
    s C= 0\,, ~~~~ S C_\mu = \partial_\mu C\,, ~~~~~
    s B_{\mu\nu} = \partial_{[\mu} C_{\nu]}\,, ~~~~~
    s \Gamma_{\mu\nu\rho}^\star = H_{\mu\nu\rho}^-
\end{eqnarray}
where $H_{\mu\nu\rho}^-$ is the anti-self-dual part of the field strength $H$. $\Gamma_{\mu\nu\rho}^\star$ is the only antifield needed to impose the self-duality condition at the level of the cohomology described by the partition function \eqref{2FC}. 
Note that the BRST symmetry is nilpotent since $H^-$ is BRST invariant. As above, the 
antifields $B^*_{\mu\nu}, C^*_\mu, C^*$ are no longer needed, since the self-duality condition imposes the equations of motion. 

We can write the last factor as follows
\begin{eqnarray}
    \label{2FE}
    \frac{(1 - 5 t + 10 t^2)}{1+t}   = 1 - \frac{6 t}{1+t} +  \frac{10 t^2}{1+t}
\end{eqnarray}
which has the following interpretation: the denominator $(1+t)$ always halves the DOFs, this implies that $H$ (which has $20$ DOF's since it is the field strength of a $2$-form) contains only the self-dual part and this gives the coefficient $10$. Since the on-shell condition is imposed by self-duality we do not need the anti-ghost. Therefore, also for ghost $C_\mu$, which, in principle should have 6 DOF's, 
  only 3 are indeed needed. Finally, we need only one ghost-for-ghost which is described by the first term.\footnote{As shown in the Maxwell case, there is yet again a direct way of deriving the partition function from the same world line model. This time, however, it is in the $R=2$ subsector of the theory. Yet again, the spectrum is much bigger then the minimal BV content, but all the extra fields imposed on us by the structure of the $N=2$ world line theory are auxiliary and therefore cancel out from the partition function.}

\subsection{Gravity}

We recall that Einstein gravity can be formulated in two equivalent ways: metric formulation with a metric and the corresponding invariances, or with Cartan formulation with a vielbein and spin-connection with their invariances. At the level of counting the fundamental DOF (pertubatively, at the quadratic level) they give the same answer and here we present what are the implication on the partition functions constructed with our scheme. Let us start from the metric formulation: in that case we have a symmetric tensor $g$, a ghost vector $\xi$ parametrizing the diffeomorphims in $D=4$, the 
antifield of the metric $g^*$ and finally the antifield $\xi^*$ of the ghost $\xi$. Putting 
all these data together we end up with the expression
\begin{eqnarray}
    \label{GRA}
    \mathbb{P}_{metric}(t) = \frac{ 4 - 10 t + 10 t^3 - 4 t^4 }{(1-t)^4} = \frac{(1-t^2)}{(1-t)^4} ( 4 (1 - t^2) - 10 t) 
\end{eqnarray}
Again by factoring out $(1-t^2)/(1-t)^4$, we can recognize the two physical degrees of freedom of the perturbative graviton on flat space. 

Let us consider the other formulation: Cartan formalism. In that case we still have diffeomorphism ghost $\xi$, but we have the Lorentz ghost $\Lambda^{[ab]}$ which carry 6 DOF ($a,b=0, \dots 3$). In order to keep track of the difference between the two approaches we introduce a second parameter $s$ as a secondary ghost number assigned to additional fields. Then, we have the vielbein $e^a$ and the spin-connection $\omega^{[ab]}$ which are 1-forms. The spin connection carries also the charge $s$. Then we have the antifields $e^*_a, \omega^*_{[ab]}$ and finally the antifields for the ghosts. We can collect all these data in the expression 
\begin{eqnarray}
    \label{GRB}
     \mathbb{P}_{Cartan}(t) = \frac{ (4 + 6 s) - (4 + 6 s) 4t + (4 + \frac{6}{s}) 4t^3 - \left(4 +\frac{6}{s}\right) t^4 }{(1-t)^4}
\end{eqnarray}
Notice that we assigned the charge $1/s$ both to the antifield to 
$\omega^{[ab]}$ and to the antifield to the Lorenzt ghost. This is needed in order to respect the anti-bracket since scaling the Lorentz ghost by $s$ implies the scaling of its antifield $\Lambda^*$ oppositely. 

In order to reconcile the two expressions we notice the following. 
The ghost $\Lambda^{[ab]}$ are needed to remove the DOF from the vielbein. That's reduces the DOF of $e^a$ from 16 to 10. However, since the BRST ghost $\Lambda^{[ab]}$ appear linearly in the transformation of the vielbein, in order to effectively cancel those DOF in the partition function we have to chose $s = t$. In this way we get 
\begin{eqnarray}
    \label{GRBA}
     \mathbb{P}_{Cartan}(t) = \frac{ (4 + 6 t) - (4 + 6 t) 4t + (4 + \frac{6}{t}) 4t^3 - \left(4 +\frac{6}{t}\right) t^4 }{(1-t)^4} = 
     \frac{ 4 - 10 t + 10 t^3 - 4 t^4 }{(1-t)^4}
\end{eqnarray}
and the interpretation is the following. The $24 t^2$ constraints coming from the BRST variation of the antifields $\omega^*$ imposes the torsion condition which can be solved by fixing $\omega$ in terms of $e^a$. Therefore there is a cancellation $ - 24 t^2 + 24 t^2 =0$. The BRST ghost $\Lambda$ has now the right powers of $t$ to cancel $6$ DOF from the vielbein and contextually there are $6$ antifeld imposing the condition of the equations of motion 
$16 t^3- 6 t^3  = 10 t^3$. Then, finally the two partition function for $s=t$ do coincide. 

\section{Supersymmetric Models}
In this section of the paper we turn our attention to models with rigid supersymmetry. As the partition function counts the on shell degrees of freedom, as an imprint of SUSY we will see the matching of fermionic and bosonic parts of the spectrum. The examples on which we show this matching are Wess-Zumino multiplet, $N=1,4$ super Maxwell and $D=4, N=1$ supergravity.  

In several models, the supersymmetry algebra closes on fermionic equations of motion (open algebra) and that prevents an off-shell description, in some cases, however the introduction of auxiliary fields may solve those issues. It is shown in the forthcoming sections how to read the auxiliary fields in the partition functions. This will help only for the first issue, namely to achieve the off-shell matching, but it does not solve the 
issue of supersymmetry algebra closure.

\subsection{WZ multiplet}
We now proceed to discussion of a first supersymmetric model of this section, Wes-Zumino model.
The WZ multiplet is formed by a complex scalar $\phi$, a Weyl spinor $\psi$ and an 
auxiliary fields $F$. The partition function reads 
\begin{eqnarray}
    \label{WZA}
     \mathbb{P}_{WZ}(t) = \frac{ - 2 t  + 4 t^{3/2} - 4 t^{5/2} + 2 t^3}{(1-t)^4} = \frac{(1- t^2)}{(1-t)^4} \left( - 2t  + \frac{4 t^{3/2}}{1+t}\right)   
\end{eqnarray}
Factoring out $\frac{(1- t^2)}{(1-t)^4}$, we are left with the expression 
\begin{eqnarray}
    \label{WZB}
    \left( - 2t  + \frac{4 t^{3/2}}{1+t}\right)
\end{eqnarray}
and in the limit $t \rightarrow 1$ it reduces to $-2 + 2 =0$, which is the on-shell matching of WZ DOFs. The first contribution to \eqref{WZB}, namely $-2 t$ is due to the scalars which have dimension one for D=4 theory, and the second piece if the contribution of the fermions. 
The sign is reversed, the coefficient is 4, because of two complex DOFs, the scale is $3/2$ as for WZ model in $D=4$ and the denominator halves the DOF going on-shell, which is what the Dirac equation does to physical DOF. 

There is another possible interpretation of \eqref{WZA} by means of BV fields. The first two terms correspond to 
the scalar $\phi$ and to the fermionc $\psi_\alpha$, the 
third term $4 t^{5/2}$ stands for the antifield $\psi^{*, \dot\alpha}$ which can be considered as the antifield of $\psi_\alpha$. 
Its sign is opposite to the previous term $- 4 t^{3/2}$, which implies that it is bosonic field, the scale $t^{5/2}$, correctly matches with its scale dimension. The last term, namely $2 t^2$, stands for the antifield $\phi^*$ of the scalar. It is complex, it has the correct scale dimension and has opposite statistic with respect to the scalars $-2 t$. The numerator denotes all possible derivatives of those fields. 
The BRST symmetry relates them is simply 
\begin{eqnarray}
    \label{WZC}
    s \phi = 0\,, ~~~~~
    s \phi^* = \partial^2 \phi + \dots\,, ~~~~ 
    s \psi_\alpha = 0\,, ~~~~
    s \psi^{*, \dot\alpha} = \partial^{\dot\alpha \beta} \psi_\beta + \dots\,. 
\end{eqnarray}
where the dots indicate we are neglecting the additional terms of the EOM for $\phi$ and $\psi_\alpha$.  The BRST differential is nilpotent. Note that by going from the first expression in \eqref{WZA} to the second one, we have put in evidence the factor $(1-t^2)$ which can be interpret as the on-shell condition (notice that it has the correct scale for the Laplacian $\partial^2$).  It means that we have removed the antifields and we have the theory on-shell (this is essentially the Green's function construction of BV formalism with a gauge fixing).    

Note that at the classical level, the presence of the antifields guarantees that we can deal with open algebras without auxiliary fields since their variation give the equations of motion needed to close the algebra (see later for an algebraic discussion on this point). Nonetheless, at the quantum level (the BV formalism works only at the classical level since the BV fields are only sources for quantum operators and they are not propagating DOF's) we need the auxiliary fields to go off-shell, the write an action and to construct the propagators of dynamical fields. We can modify the above expression as follows 
\begin{eqnarray}
     \label{WZD}
     \mathbb{P}_{WZ}(t) &=& \frac{ - 2 t  + 4 t^{3/2} - 2 t^2 + 2 t^2 - 4 t^{5/2} + 2 t^3}{(1-t)^4}    = \nonumber \\ 
&=&     
      \frac{ - 2 t  + 4 t^{3/2} - 2 t^2}{(1-t)^4} + 
       \frac{2 t^2 - 4 t^{5/2} + 2 t^3}{(1-t)^4} 
\end{eqnarray}
where we added and subtracted the term $- 2 t^2$. This stands for the auxiliary fields $F$, it has the correct sign, the multiplicity and the scale dimension. Then we see the polynomial 
$- 2 t  + 4 t^{3/2} - 2 t^2$ vanishes in the limit $t\rightarrow 1$ beacuse of the off-shell matching of DOFs. The second term in the last expression of \eqref{WZD} describes all antifields to fields and to the auxiliary fields, usually denoted by $F^\star$. 

Note that the first term of \eqref{WZD} can be recast in the following form 
\begin{eqnarray}
    \label{WZE}
    \mathbb{P}_{WZ, fields}(t) &=& 
      \frac{ - 2 t  + 4 t^{3/2} - 2 t^2}{(1-t)^4} = 
      - 2 t \frac{ (1 - \sqrt{t})^2}{(1-t)^4}   
       \end{eqnarray}
This expression has also a very intuitive (worldline) description: 
the factor $(1 - \sqrt{t})^2$ is the partition function of an anticommuting variable $\theta_\alpha$ where $\alpha =1,2$, which scale as $\sqrt{t}$, which is the correct engineering dimension of the target space super coordinates. Infact the superfield $\Phi$ built with this coordinates is 
\begin{eqnarray}
    \label{WZF}
    \Phi(x, \theta) = \phi(x) + \psi_\alpha(x) \theta^\alpha + F(x) \frac{\epsilon_{\alpha\beta}}{2} \theta^\alpha \theta^\beta
\end{eqnarray}
where $\phi(x), \psi_\alpha(x)$ and $F(x)$ are the component fields and these provide the coefficents of the polynomial $(1 - \sqrt{t})^2$. 
Then the factor $(1-t)^4$ in the denominator stands for the bosonic components $x^\mu$ of target space coordinates. So, we can conclude that the partition function of the WZ multiplet can be obtained by considering a Green-Schwarz superparticle whose target space 
is D=4 chiral superspace. It remains to explain the factor 
$- 2t$. The scale $t$ is appropriate for the interpretation, the coefficient $2$ stands for complex DOF, the negative sign is needed since in the plethystic exponential indicates that those are bosonic superfields and therefore they provide positive coefficients in the expansion. 

\subsection{$N=1$ Super Maxwell}

Let consider the Super-Maxwell case $N=1$ in $D=4$. The partition function is 
\begin{eqnarray}
    \label{SMA}
     \mathbb{P}_{SM}(t) = \frac{ 1 - 4 t  + 4 t^{3/2} - 4 t^{5/2} + 4 t^3 - t^4}{(1-t)^4} = \frac{(1- t^2)}{(1-t)^4} 
     \left( 1 + t^4 - 4t  + \frac{4 t^{3/2}}{1+t}\right)   
\end{eqnarray}
Notice that switching off the $t^{3/2}, t^{5/2}$ terms, it reduces to the Maxwell case \eqref{SDA}. The additional contributions 
are the fermions DOF's. In the last expression, we have factor 
out the on-shell condition and the derivatives and we are left with the expression 
\begin{eqnarray}
    \label{SMB}
    \left( 1 + t^4 - 4t  + \frac{4 t^{3/2}}{1+t}\right) 
\end{eqnarray}
with the following interpretation: $(1+ t^2)$ are the ghost and the antighost field (as is well known, one can assign zero scaling dimension to the ghost $C$ 
and the complement to the antighost $B$). Again being fermionic DOF, the sign is positive. The term $- 4 t$ stands for the gauge bosons $A_\mu$, the dimension, the sign and the coefficient follows the rules we discussed. The forth piece describes the fermions: the have scale dimesnions $t^{3/2}$, four are DOF's, the sign is positive and the factor $1/(1+t)$ halves the DOF's on-shell (in the limit $t\rightarrow 1$). Note that in that limit 
there is a matching of DOF's, which is the well-known on-shell mathcing. 

Let us check the second interpretation. We notice that 
number can be interpret as follows: the first constant $1$ stands for the ghost field $C$ (it has vanishing dimension, it is a fermion and it is a Lorentz scalar), the next one: $- 4 t$ stands for the gauge field $A_\mu$: four DOF's, bosonic and scale dimension one. The next one is the fermion $\psi_\alpha, \bar\psi_{\dot\alpha}$. Then we have $ 4 t^{5/2}$ has to be compared with the antifield to the fermions 
$\psi^*_\alpha, \bar\psi^*_{\dot\alpha}$ matching again the quantum numbers. The term $+ 4 t^3$ corresponds to the antifield of the gauge field $A^*\mu$ and, finally, $- t^4$ reads as the antifield of the ghost field $C^*$. The BRST symmetry relating those fields contains the equations of motion as variations of the antifield and the usual gauge symmetry. The factor $1/(1-t)^4$ are the derivatives acting on all of these fields. 

\subsubsection{Auxiliary Fields}

In order to study the auxiliary fields of this model we can perform the same analisys as for WZ model. 
We notice that we can also add and subtract a term as $-t^2$ to the numerator of \eqref{SMA} to get 
\begin{eqnarray}
    \label{SMC}
     \mathbb{P}_{SM}(t) &=& \frac{ 1 - 4 t  + 4 t^{3/2} - t^2 + t^2- 4 t^{5/2} + 4 t^3 - t^4}{(1-t)^4}  \nonumber \\
      &=& \frac{ 1 - 4 t  + 4 t^{3/2} - t^2}{(1-t)^4} + 
       \frac{t^2- 4 t^{5/2} + 4 t^3 - t^4}{(1-t)^4} 
\end{eqnarray}
where we have separated the expression in two parts that we are calling $\mathbb{P}_{SM, fields}(t)$ and $\mathbb{P}_{SM, antif}(t)$. The first one represents the field part of the partition function where we added the term $-t^2$. This can be read as the auxiliary field $D$ needed in $N=1$ $D=4$ case. The polynomial 
\begin{eqnarray}
    \label{SMD}
    p(t) = 1 - 4 t  + 4 t^{3/2} - t^2
\end{eqnarray}
has the property to vanish at $t =1$, and that corresponds to 
the off-shell matching (we have not extract the coefficient $(1-t^2)$) for super Maxwell theory. The additional piece of the partition function collects the antifield together with $D^*$, the antifield of the auxiliary field $D$. 

In order to find out if there is superparticle model which might have such partition function, we make the following ansatz: we need at least 4 supercoordinates $\theta^\alpha$  and $\bar \theta^{\dot\alpha}$ and their partition function is $(1- \sqrt{t})^4$ as above. Now, we use the simple identity  
\begin{eqnarray}
    \label{SME}
    (1 - \sqrt{t})^4 = 2 (1 - \sqrt{t})^2 - (1 - 4 t + 4 t^{3/2} - t^2) 
\end{eqnarray}
to relate it to the polynomial in \eqref{SMD}. Note that there is a contribution of the piece $2 (1 - \sqrt{t})^2$, but that corresponds (as seen as above) the a chiral multiplet. Indeed, the super Maxwell multiplet is obtained by removing from a real  
general superfield $\Phi(x, \theta, \bar \theta)$ a chiral multiplet unessential for physics by means of the WZ gauge. Therefore, provided this interpretation, we can conclude that the 
partition function of $N=1$ $D=4$ can be constructed by means a GS superparticle with $D=4$ complete superspace $(x, \theta, \bar \theta)$, or equivalenty with $N=2$ supersymmetric particle as in \cite{Boffo:2024lwd}. 

\subsection{Self-Dual $N=1$ Super Maxwell}

From the N=1 super Maxwell case we can extract the self-dual counterpart. 
As been done in the previous section, we add and subtract some contribution as follows 
\begin{eqnarray}
    \label{SDSMA}
     \mathbb{P}_{SM}(t) &=& \frac{ 1 - 4 t  + 4 t^{3/2} + 3 t^2 - 2 t^{5/2}  - 3 t^2- 2 t^{5/2} + 4 t^3 - t^4}{(1-t)^4} \nonumber \\  
     & \longrightarrow&       \mathbb{P}_{SDSM}(t) = \frac{ 1 - 4 t  + 2 t^{3/2} + 3 t^2 - 2 t^{5/2}}{(1-t)^4} 
     \end{eqnarray}
where the antifield part of the partition function is removed and we are left with the self/anti-self part of the 
partition function. Notice that we have add the contribution $+ 3 t^2$ as in the non-susy case, 
we keep half of the fermion wth  
half of the antifield $\psi^*$ of the fermion to implement its equation of motion (in the fermionic case the 
chirality condition is not sufficient to impose the equations of motion) 
Again factoring out $(1-t^2)/(1-t)^4$ we are left with 
\begin{eqnarray}
    \label{SDSMB}
    \left(
    1 - \frac{4 t}{(1+t)} + \frac{2t^{3/2}}{(1+t)}
    \right) 
\end{eqnarray}
and in the limit $t \rightarrow 1$, there is a matching between the self-dual photon DOF and the 
on-shell chiral fermion. The denominators $1/(1+t)$ have the duty to halve the DOF's. 

Let us return to the discussion on antifields and their BRST symmetry. As we have already noticed the term $3 t^3$ have been understood as the presence of an antifield, which is a 2-form whose variation is the self-dual field strength. On the other hand, we 
have to take into account  the term $- 2 t^{5/2}$, this are the antifields for the fermion $\psi_\alpha$ needed to implement the equations of motion, we finally have 
\begin{eqnarray}
    \label{SDSMBA}
&&s C =0\,, ~~~~~~
s A_\mu = \partial_\mu C\,, ~~~~~~
s \Sigma^{+}_{\mu\nu} = F^{+}_{\mu\nu} \,, ~~~~~~\nonumber \\
&&s \psi_\alpha = 0\,,  ~~~~~~~~~
s \psi^*_{\dot \alpha} = \partial_{\dot \alpha \alpha} \psi^\alpha 
\end{eqnarray}
In the case of self-dual Maxwell, removing the anti-chiral part of the fermion, there is no way to write the Dirac action \cite{Barbagallo:2022kbt}, however the equations of motion are easily implemented at the level of the cohomology by the variation of the antifields. Note again the asymmetry between fields and antifields and that prevents us from getting the BRST transformations from a master action $S$ leading to the antibracket. Again, the partition function seems to suggest an asymmetrical treatment of fields and antifields.

\subsection{$N=4$ Super Maxwell}

Let us come to the most interesting case: the $N=4, D=4$ super Maxwell.\footnote{Everything in the present section works also for $N=2, D=4$ super Maxwell theory with appropriate changes and we do not discuss here.} In that case the partition function is 
\begin{eqnarray}
    \label{SM4A}
     \mathbb{P}_{SM4}(t) = \frac{ 1 - 10 t  + 16 t^{3/2} - 16 t^{5/2} + 10 t^3 - t^4}{(1-t)^4} = \frac{(1- t^2)}{(1-t)^4} 
     \left( 1 + t^2 - 10t  + \frac{16 t^{3/2}}{1+t}\right)   
\end{eqnarray}
where again we have the twofold interpretations and we are not repeating it. Note that the treatment of the super Maxwell with antifields has been done and, at the classical level, it provide a way out for open supersymmetry algebra without introducing auxuliary fields. 

The matching of on-shell DOF's is achieved by 
computing the limit $t\rightarrow 1$ of the last factor of the last expression leading to $-8 +8$ cancellation. Note that in the 
first sector we have $1 + t^2 - 10 t$, which correspond to 
the ghost $C$, the antighost $B$, the gauge fields $A_\mu$ and the scalar fields $\phi_{IJ}$ (six real scalar field whose scale dimension is the same as of the gauge fields). 

Let us introduce the auxiliary fields. For that we notice that we can insert in the expression a term $- 7 t^2$ and subtracting at the same time. That leads to 
\begin{eqnarray}
    \label{SM4B}
     \mathbb{P}_{SM4}(t) = \frac{ 1 - 10 t  + 16 t^{3/2} - 7 t^2}{(1-t)^4} + 
     \frac{7 t^2 - 16 t^{5/2} + 10 t^3 - t^4}{(1-t)^4} 
\end{eqnarray}
where we separated again the expression into field and antifield part. The additional term $- 7 t^2$ is needed in order to have the off-shell matching and the coefficient stands for one auxiliary field $D$ for the gauge multiplet plus a set of $6$ auxiliary fields $F^{IJ}$ for the scalars. In that case, we can achieve the matching off-shell of DOF's. 

The number $7$ is not a choice, but we can recall that the $N=4$ on-shell spectrum can be view as a gauge boson plus 6 scalars \cite{Berkovits:1993hx,Baulieu:2007ew,Pestun:2009nn}. Therefore, we should be able to get it also from the partition-function point of view. This can be done by summing the contribution of $N=1$ super Maxwell computed in \eqref{SMD} 
to 6 WZ multiplets \eqref{WZD}
\begin{eqnarray}
    \label{SM4C}
    (1 - 4 t + 4 t^{3/2} - t^2 ) + \left( - 6 t (1 - 2 \sqrt{t} + t) \right) =  1 - 10 t  + 16 t^{3/2} - 7 t^2
\end{eqnarray}
leading to the correct expression. The plus takes into account  also the representations of the fields, a more refined version of 
\eqref{SM4C} can be exploited using Young tableaux. 

\subsection{D=6 N=1,2 Self-Dual Supersymmetric 2-form}
As a next example we consider the case of $2$-form in $D=6$ with $N=1,2$ supersymmetric theory. For higher forms there is no derivation of the partition functions using the 
algebraic techniques (see for example \cite{ duboisviolette2024quadraticalgebrasassociatedexterior}) and therefore we 
use the knowledge of the previous sections to construct the partition function as follows. We start from the on-shell degrees of freedom which are 3 DOF from the self-dual 2-form, 4 DOF for the fermonic DOF and 1 from a scalar completing the multiplet (see \cite{Howe:1983fr}) then we have 
\begin{eqnarray}
    \label{S2FA}
    \mathbb{P}_{2FN1}(t) 
    &=& \frac{(1- t^2)}{(1-t)^6} \Big(1 - \frac{6t}{(1+t)} + \frac{10 t^2}{(1+t)} + t^2 - \frac{8 t^{5/2}}{(1+t)}\Big)   
\end{eqnarray}
where we added the scalar field $t^2$ and the fermions. Note that 
in order to achieve the correct DOF we have provided the correct signs. Taking the limit $t\rightarrow 1$ this gives $+4 - 4 =0$ matching of the on-shell DOF. Expanding the overall factors 
we finally get the final expression 
\begin{eqnarray}
    \label{S2FB}
    \mathbb{P}_{2FN1}(t) = \frac{1 - 6 t + 16 t^2 - 8 t^{5/2} - 10 t^3 + 8 t^{7/2} - t^4}{(1-t)^6}
\end{eqnarray}
In terms of BV fields, we find that the first terms $(1,-6, 15+1, -8)$ are 
the physical fields with the ghost and the ghost-for-ghost, while the additional terms $(-10, 8, -1)$ are the antifields. Note that $-10 t^3$ represent the antifields DOF for the self-duality (as discussed above), $+ 8 t^{7/2}$ the antifield for the fermions (implementing the equations of motion) and finally $- t^4$ is the antifield for the scalar implementing its equation of motion. 

For the case N=2 we have 
\begin{eqnarray}
    \label{S2FBA}
    \mathbb{P}_{2FN2}(t) = \frac{1 - 6 t + 18 t^2 - 16 t^{5/2} - 10 t^3 + 16 t^{7/2} - 3 t^4}{(1-t)^6}
\end{eqnarray}
representing the multiplet of N=2 self-dual form in D=6. 

\subsection{N=1 D=4 Supergravity}
In the last section we depart from rigid supersymmetric models and give an analysis of linearized  supergravity partition function 
    \begin{eqnarray}
    \label{SGA}
     \mathbb{P}_{SG}(t) &=& \frac{ 4 - 4 \sqrt{t} - 10 t  + 16 t^{3/2} - 16 t^{5/2} + 10 t^3 + 4 t^{7/2}- 4 t^4}{(1-t)^4} \nonumber \\
     &=& 
     \frac{(1- t^2)}{(1-t)^4} 
     \left( 4(1 + t^2) - 10t  + \frac{\sqrt{t}(16 t - 4 (1+ t + t^2)}{1+t}\right)   
\end{eqnarray}
The interpretation of the last factor is the following 
$4(1+ t^2)$ are the ghost and antighost of diffeomorphisms, 
and $- 10 t$ are the independent components of the metric tensor. 
In the second term we have $16 t^{3/2}$ are the components of the gravitino and $- 4(1 + t + t^2)$ are the supersymmetric ghost and the on-shell traceless-condition. The minus sign is needed since they are commuting ghosts. Taking the limit $t \rightarrow 1$, one can observe the on-shell matching  condition for $-2 + 2$, of the two graviton states and those of gravitino. 

\section{Dualities} \label{sec5.}

All the models -- aside of the ones wich are cut down by self duality -- admit a duality at the level of the partition function. Here we list these models alongside with the duality transformation $t \rightarrow 1/t$ 
\begin{center}
\begin{tabular}{|c|c|c|}
\hline
Fields & Partition Function & Duality  \\
\hline
MX & $\mathbb{P}_{MX}(t) = \frac{1 - 4 t + 4 t^3 - t^4 }{(1-t)^4}$ & $\mathbb{P}_{MX}(1/t) = - \mathbb{P}_{MX}(t)$ \\
\hline
2-FORM & $\mathbb{P}_{2FO}(t) =\frac{1 - 6 t + 15 t^2 - 15 t^4 + 6 t^5 - t^6}{(1-t)^6} $ &  $\mathbb{P}_{2FO}(1/t) = - \mathbb{P}_{2FO}(t)$ \\
\hline
WZ-MULT & $\mathbb{P}_{WZ}(t) =\frac{ - 2 t  + 4 t^{3/2} - 4 t^{5/2} + 2 t^3}{(1-t)^4} $ & $\mathbb{P}_{WZ}(1/t) = - \mathbb{P}_{WZ}(t)$ \\
\hline
N=1 SMX & $\mathbb{P}_{SMX-1}(t) = \frac{ 1 - 4 t  + 4 t^{3/2} - 4 t^{5/2} + 4 t^3 - t^4}{(1-t)^4}   $ & $\mathbb{P}_{SMX-1}(1/t) = - \mathbb{P}_{SMX-1}(t)$ \\
\hline
N=4 SMX & $ \mathbb{P}_{SMX-4}(t) = \frac{ 1 - 10 t  + 16 t^{3/2} - 16 t^{5/2} + 10 t^3 - t^4}{(1-t)^4}
$ & $\mathbb{P}_{SMX-4}(1/t) = - \mathbb{P}_{SMX-4}(t)$ \\
\hline
N=1 SG & $  \mathbb{P}_{SG}(t) = \frac{ 4 - 4 t^{1/2} - 10 t  + 16 t^{3/2} - 16 t^{5/2} + 10 t^3 + 4 t^{7/2}- 4 t^4}{(1-t)^4}$ &
$\mathbb{P}_{SG}(1/t) = - \mathbb{P}_{SG}(t)$ \\
\hline
\end{tabular}
\end{center}
\label{defaultA}
Physical interpretation of this duality is just a relation of fields to antifields. The minus sign is the change of the statistic going from one to each other.

\section{Conformal field theories}

In the case of Maxwell and of higher forms in sec. 3.1, we have show how the partition function can be computed from the worldvolume fields starting from a simple worldline model. One might wonder whether this can be done in any case studied in the present paper. We do not have any conclusive answer, but we can observe that 
if we can reproduce those expressions for a set of 2d fields, with a $(1,0)$ CFT of the fermionic type $b_I,c_I$ or of the bosonic type $\beta_I,\gamma_I$ by computing the partition function of their zero modes. 
The partition functions discussed above allows us to extract the dimensions of the representation for the 2d fields and in terms of which we can compute some of the CFT parameters that characterize the model. 

Once we consider $(1,0)$ CFT systems, we can establish their conformal algebra: 
the energy momentum tensor $T(z), \bar T(\bar z)$ and the 
current $J(z), \bar J(\bar z)$ which are spin 2 and spin 1 fields, respectively. Computing the algebra we expect the general form
\begin{eqnarray}
    \label{COA}
    T(z) T(w) &\sim & \frac{c_{Vir}}{(z-w)^4} + 
    \frac{T(w)}{(z-w)^2} + \frac{\partial T}{(z-w)} + \dots 
    \nonumber \\
    T(z) J(w) &\sim & \frac{c_g}{(z-w)^3} + 
    \frac{J(w)}{(z-w)^2} + \frac{\partial J}{(z-w)} + \dots 
    \nonumber \\
    J(z) J(w) &\sim & \frac{a_g}{(z-w)^2} + \dots 
\end{eqnarray}
and analogously for the anti-holomorphic sector. The 
computation of numbers $c_{Vir}, c_g, a_g$ can be computed in 
terms of the representation of the fields, their charges and their conformal spin. This is usually very difficult since the dimension of representations grow very rapidly. Nonetheless, we can adopt a different route. 

We first compute the number of independent fields at each stage. 
This can be done by recasting the above partition functions in terms 
of infinite products 
\begin{eqnarray}
    \label{COB}
    \mathbb{P}(t) =\frac{P(t)}{Q(t)} = \prod_{n=1}^\infty (1 - t^n)^{-N_n}
\end{eqnarray}
where $N_n$ are integers. The above expression stands for the contribution to the zero mode partition function computed with conformal fileld theory techniques on the torus Riemann surface 
\cite{Berkovits:2005hy}. 
To compute those terms we replace $t \rightarrow e^x$, and we compute 
\begin{eqnarray}
    \label{COC}
    -\log\frac{P(e^x)}{Q(e^x)} = \sum_{n=1}^\infty N_n \log(1 - e^{n x}) 
\end{eqnarray}
since 
\begin{eqnarray}
    \label{COD}
    \log(1 - e^{x}) = \log(-x) +\frac{x}{2} + \sum_{g=1}^\infty \frac{B_{2g}}{2g (2g)!} x^{2g} = 
    \log(-x) + \frac{x}{2} + \frac{x^2}{24} + \dots
\end{eqnarray}
where $B_k$ are the Bernoulli numbers and then, inserting it into \eqref{COC}, 
we have 
\begin{eqnarray}
    \label{COE}
    \log(-x) \sum_n N_n + \sum_n \log(n) N_n + \frac{x}{2} \sum_n {n N_n} + 
\sum_{g=1}^\infty \frac{B_{2g}}{2g (2g)!} x^{2g} \sum_n n^{2g} N_n = -\log\frac{P(e^x)}{Q(e^x)} \nonumber 
\end{eqnarray}
and therefore we can extract the 
the sums of coefficients which can be finally identified with the central charges in \eqref{COA} as follows  
\begin{eqnarray}
    \label{COF}
    c_{Vir} = 2 \sum_n N_n\,, ~~~~
    L = \sum_n \log(n) N_n\,, ~~~~~
    a_g = - \sum_n n N_n \, ~~~~
    c_g = - \sum_n n^2 N_n \,, 
\end{eqnarray}
which are the coefficients in front of $\log(-x),1,x/2$ and $x^2/24$ in the expansion with Bernoulli numbers. 

We computed this number and we list here 
\begin{table}[htp]
\begin{center}
\begin{tabular}{|c|c|c|c|c|}
\hline
Fields & $C_{Vir}$ & $L$ & $a_g$ & $c_g$  \\
\hline
MX & $6$ & $\log (-4)$ & $0$ & $48$ \\
\hline
SD-MX & $6$ & $\log (-2)$ & $0$ & $48$ \\
\hline
2-FORM & $10$ & $\log(12)$ & $0$ & $- 5/4$\\
\hline
SD-2-FORM & $10$ & $\log(6)$ & $0$ & $- 5/4$\\
\hline
\\
\hline
WZ-MULT & $2$ & $\log(-1/2)$ & $0$ & $-5/2$  \\
\hline
N=1 SMX & $2$ & $\log(3/2)$ & $0$ & $- 7/2$ \\
\hline
N=2 SMX & $2$ & $0$ & $0$ & $- 13/2$ \\
\hline
N=4 SMX & $2$ & $\log(1/2)$ & $0$ & $1$ \\
\hline
\end{tabular}
\end{center}
\label{defaultB}
\end{table}%
We will return on this problem in the future. At the moment, we leave the question open and we provide only these observations. 

\section{Plethystic exponential}

Last step to obtain the full flagged multi word partition function is to use a pletistic exponental on the single letter word partition function, these techniques were studied for instance in \cite{Feng:2007ur}.

As mentioned above we now only consider a specialized partition function with one fugacity and cohomological degree taken into account by specializing the appropriate fugacity to $-1$. Given such a partition function in the expanded form  

\begin{eqnarray}
    \label{PLEA}
\mathbb{P}(t) = \sum_{n=1}^\infty C_n t^n
\end{eqnarray}
where $C_n$ are integer numbers, the sign is informing us about the bosonic/fermionic nature of the given space. Then we can build the multifield expression using the plethystic exponential 
\begin{eqnarray}
    \label{PLEB}
    PE[\mathbb{P}(t) x] = \prod_{n=1}^\infty (1 - x t^n)^{C_n} 
\end{eqnarray}
where the parameter $x$ counts the number the letters in each word of given fixed fugacity. Notice that \eqref{PLEA} is a series because of the multiple derivatives of the fields. Therefore, \eqref{PLEB} contains all posibile monomials with fields and their derivatives. Furthermore the signature of $C_n$ produces either appropriate fermionic/bosonic contribution.

Let us consider the simplest case of Mawxell $D=4$, the partition function is depicted in \eqref{SDA}, if we put the formula in the following form 
\begin{eqnarray}
    \label{PLEC}
    \mathbb{P}_{MX} (t) = \frac{1+t}{(1-t)^3} (1 + t^2 - 4t)
\end{eqnarray}
then we can use the identity 
$$
 \frac{1+t}{(1-t)^3} = \sum_{n=1}^\infty n^2 t^{n-1}
$$
which inserted in \eqref{PLEC} gives 
\begin{eqnarray}
    \label{PLED} 
    \mathbb{P}_{MX} (t) = 1 + 2\sum_{n \geq 2} (n^2-1) t^{n}
\end{eqnarray}
which is in the form useful to extract the Plethystic Exponential 
\begin{eqnarray}
    \label{PLEE}
PE[\mathbb{P}(t) u] &=& \frac{1}{1 - u\, t}{}\prod_{n\geq 2}\frac{1}{(1 - u\,  t^n)^{2(n^2-1)}}
\\ &=&
1 + 
\left(48 t^5+30 t^4+16 t^3+6 t^2+t\right) u \nonumber 
\\
&+&\left(1176 t^{10}+1440 t^9+1233 t^8+768 t^7+364 t^6+126 t^5+37 t^4+6 t^3+t^2\right) u^2 + O\left(u^3\right) \nonumber 
 \end{eqnarray}
 where $u$ counts the number of fields. The linear term in $u$ are the single letter operators, the second line he double letter operators, etc....

\vfill
\eject 

\section*{Acknowledgements}
We thank C.A. Cremonini, F. Morales-Morera, I. Sachs, R. Norris for several interesting discussions.

\bibliographystyle{unsrt}
\bibliography{N=1.bib}

\end{document}